\documentclass[prb,twocolumn,amsmath,amssymb,aps,superscriptaddress,eqsecnum]{revtex4-1}
\usepackage{graphicx}
\usepackage{dcolumn}
\usepackage{bm}
\begin{document}

\title{Geometry defects in Bosonic symmetry protected topological phases }

\author{Yizhi You}
\affiliation{Department of Physics and Institute for Condensed Matter Theory, University of Illinois at Urbana-Champaign, 
Illinois 61801}
\affiliation{Kavli Institute for Theoretical Physics, University of California Santa Barbara, Santa Barbara, California 93106}
\author{Yi-Zhuang You}
\affiliation{Kavli Institute for Theoretical Physics, University of California Santa Barbara, Santa Barbara, California 93106}

\date{\today}
\begin{abstract}
In this paper we focus on the interplay between geometry defects and topological properties in bosonic symmetry protected topological(SPT) phases. We start from eight copies of 3D time-reversal($\mathcal{T}$) invariant topological superconductors(TSC) on a crystal lattice. We melt the lattice by condensation of disclinations and therefore restore the rotation symmetry. Such disclination condensation procedure confines the fermion and afterwards turns the system into a 3D boson topological liquid crystal(TCL). The low energy effective theory of this crystalline-liquid transition contains a topological term inherited from the geometry axion response in TSC. In addition, we investigate the interplay between dislocation and superfluid vortex on the surface of TCL. We demonstrate that the $\mathcal{T}$ and translation invariant surface state is a double $[e\mathcal{T}m\mathcal{T}]$ state with intrinsic surface topological order. We also look into the exotic behavior of dislocation in 2D boson SPT state described by an $O(4)$ non-linear $\sigma$-model(NL$\sigma $M) with topological $\Theta$-term. By dressing the $O(4)$ vector with spiral order and gauge the symmetry, the dislocation has mutual semion statistics with the gauge flux. Further reduce the $O(4)$ NL$\sigma M$ to the Ising limit, we arrive at the Levin-Gu model with stripy modulation whose dislocation has nontrivial braiding statistics.
\end{abstract}

\maketitle

\section{Introduction and Motivation}
Throughout the past few years, much effort had been made on classification and realization of symmetry protected topological phases (SPT)\cite{wang2014classification,chen2010local,chen2012symmetry,levin2012braiding,bi2015classification}. Even the SPT states themselves are short range entangled, the surface state of SPT phase can exhibit topological order in an anomalous way\cite{metlitski2013bosonic,wang2014braiding,bi2015classification,burnell2014exactly,wang2014interacting,metlitski2014interaction,jian2014layer,fidkowski2013non,chen2014symmetry,you2014symmetry,xu2013wave,jiang2014generalized}. In addition, despite lack of anyonic quasiparticle excitation in SPT phase, once we gauge the symmetry, the gauge flux contains nontrivial braiding statistics and symmetry gauging procedure provides a straightforward way to bridge SPT and topological order\cite{levin2012braiding,xu2013wave,senthil2013integer,bi2014anyon,wang2015field}.

The blossom of SPT states provides a rich platform to study the interplay between geometry and topological degree of freedom. In 2D Chern insulator coupled to gravity, there exists a gravitational Chern-Simons response as a consequence of thermal Hall conductance\cite{you2014theory,you2013field,cho2014geometry,abanov2014electromagnetic,gromov2015framing,hoyos2012hall,hoyos2014hall}; in 3D time reversal invariant topological insulator/superconductor coupled to gravity, one expects a gravitational $\theta$ term as a signature of Gravitational Witten effect and Gravitoelectromagnetism \cite{wang2011topological,wang2011topological,ryu2012electromagnetic}.  Beyond these exotic gravitational responses, lattice defects in some topological lattice models contain non-abelian braiding statistics and therefore change the topological nature of the system.\cite{jian2014layer,you2013synthetic,barkeshli2013theory,barkeshli2013twist,bombin2010topological}
In topological Crystalline insulators/superconductors, lattice defect can carries Majorana zero mode, which propose a new way to create and manipulate the Majorana fermions.\cite{benalcazar2014classification,cho2015condensation,teo2013existence,hughes2014Majorana,gopalakrishnan2013disclination,ran2009one,barkeshli2013theory}.

In this paper, we intend to step forward the study on the intertwined phenomenon between geometry defect and topological matter. First, we start from eight copies of time-reversal($\mathcal{T}$) invariant topological superconductor(TSC)\cite{wang2014interacting,metlitski2014interaction} in 3D on a crystalline lattice. By lattice melting and disclination condensation, one can restore the translation/rotation symmetry and meanwhile system undergoes a crystalline-liquid transition\cite{cho2015condensation,gopalakrishnan2013disclination}. As the fermions view the disclinations as a $\pi$ flux of the spin connection\cite{bi2014bridging}, the proliferation of disclinations would therefore confine the fermion degree of freedom. Consequently, disclination condensation drives the crystalline lattice into a boson isotropic Topological Liquid Crystal(TLC). The bulk of such boson isotropic liquid crystal is short range entangled, but the surface can exhibit novel surface topological order(STO) and the translational/rotation symmetry plays an important role.

In addition to the surface topological order via lattice melting, the interplay between geometry defect and STO in $\mathcal{T}$ invariant TSC provide a new way to identified the $Z_{16}$\cite{wang2014interacting,metlitski2014interaction,chen2014symmetry} classification scheme. As is illustrated in several pioneer works\cite{wang2011topological,ryu2012electromagnetic,gu2015multi}, the geometry(gravitational) axion term alone is only invariant mod $2\pi$ and the geometric response is not enough to distinguish the $Z_{16}$ classification of interacting TSC. We would show in the rest of the paper that the braiding statistics of the geometry defects ($e.g.$ dislocation, disclination) on the surface can identify the 16 fold way of 3D $\mathcal{T}$ invariant TSC.

Motivated by the exotic properties of geometry defects at the surface of Topological Liquid Crystal(TLC), we also investigate the topological characterization of dislocations in 2D boson SPT phase. The associate boson SPT state is described by the O(4) non-linear $\sigma$-model(NL$\sigma$M) with topological  $\Theta$ term for $\Theta=2\pi$\cite{xu2013wave,bi2015classification}. By breaking O(4) symmetry to $U(1)\times U(1)$ and dressing each $U(1)$ rotor with a spiral order, the dislocations manifest mutual semion statistics with the gauge flux(after gauging the symmetry). We further reduce $U(1)\times U(1)$ symmetry to $Z_2$ and study the stripy modulated Levin-Gu model where dislocation has nontrivial braiding with the $Z_2$ flux\cite{levin2012braiding}.

The rest of this paper is organized as follows. In section~\ref{tlc}, we investigate the field theory description of the 3D topological liquid crystal(TLC). In section~\ref{surfaceTLC}, we studied the several possible surface state of TLC. In section~\ref{gu}, we study a similar case in 2D boson SPT and explore the novel property of dislocation in such system.

\section{Crystalline-liquid transition from disclination condensation in 3D TSC}
\label{tlc}
The classification of 3D $\mathcal{T}$ invariant topological superconductor(symmetry class DIII) as a low energy effective theory of $^{3}$He superfluid B phase\cite{leggett1975theoretical} is well studied. In the non-interacting level, the classification is $Z$ while in the interacting level the classification reduces to $Z_{16}$\cite{wang2014interacting,metlitski2014interaction,chen2014symmetry,fidkowski2013non}. For eight copies of 3D TSC, interaction can drive the system into a boson state where the fermion degree of freedom is absent in low energy spectrum \cite{xu2013wave,bi2015classification,wang2014classification}. The gapped $\mathcal{T}$  symmetric surface of such 3D TSC is known as $e\mathcal{T}m\mathcal{T}$ state where $e$ and $m$ particle are mutual semions, each carries a Kramers doublet\cite{wang2014interacting,metlitski2014interaction,chen2014symmetry}.

In this paper, we put the 3D TSC on a crystalline lattice and proliferate the disclination/dislocation to melt the lattice\cite{cho2015condensation,fisher1979defects,gopalakrishnan2013disclination,halperin1978theory,strandburg1988two,kleinert2004nematic}.  As we are mainly interested in the disclination behavior in boson symmetry protected topological phases, we start with eight copies of 3D TSC, which is the minimal copy to acquire a boson SPT state from interacting TSC\cite{xu2013wave,bi2015classification}.
During the lattice melting procedure, the lattice degree of freedom, which can be written in terms of background vielbein becomes dynamical. The translation symmetry is restored if one proliferates and condenses the dislocation, driving the system into a nematic crystal. Further restoring the rotation symmetry can be achieved by disclination loop condensation. 

The $\mathcal{T}$ invariant topological superconductor contains spinful fermions with triplet pairing, whose low energy effective theory is,
\begin{align} 
&H=\Psi^{\dagger} (\epsilon(k) \sigma_z+|\Delta| \sigma_x k_i \tau_i) \Psi \nonumber\\
&\Psi=(c_{k,\uparrow},c_{k,\downarrow},c^{\dagger}_{-k,\downarrow},-c^{\dagger}_{-k,\uparrow})
\end{align}

In a Lorentz invariant system whose geometry is torsion free, the coupling between vielbeins and Dirac fermion is well studied. Since Dirac fermion has diffeomorphism covariance, when we rotate the frame locally, the Dirac theory $\bar{\Psi} \gamma_{\mu} \partial_{\mu} \Psi $ is invariant in the flat tangent space. After a Lorentz transformation, the spinor transforms as $\Psi \rightarrow  \exp[ i \frac{\theta^{ab} \gamma^{ab}}{2}]\Psi$. The Dirac theory in curved space is therefore $\bar{\Psi} \gamma_{\mu} e^{\mu}_{\nu} (\partial_{\nu} +\frac{i}{2}\omega^{ab}_{\nu}\gamma^{ab})\Psi $. The vielbeins both couple with the torsion tensor and appear in the spin connection in the covariant derivative. For the topological superconductor, similar argument also applies so the disclination field couples with the superconducting fermion in the same way,
\begin{align} 
&H=\Psi^{\dagger} (\epsilon(k) \sigma_z+|\Delta| \sigma_x (k_i+\frac{\omega^{ab}_{i}}{2} \sigma^{ab})e^{i}_{j} \tau_j) \Psi \nonumber\\
&\omega^{ab}_{i}= e^{c a} \nabla_i e^b_{c}
\end{align}
Here we use the  Einstein summation convention and $a,b,c,i,j$ runs over the spatial index.
$d\omega^{ab}=-d\omega^{ba}$ is the disclination density in the $a$-$b$ plane. To simplified the expression, we would use the notation $\omega^{[ab]}$ in our future content where $\omega^{[ab]}=\omega^{ab}=-\omega^{ba}$. 

\begin{figure}[h]
    \centering
    \includegraphics[width=0.35\textwidth]{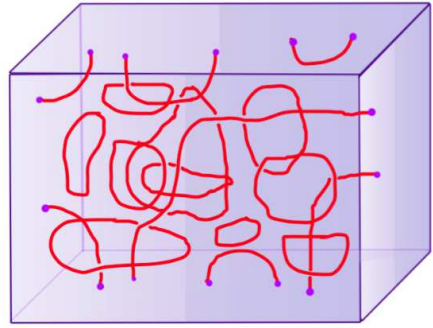}
    \caption{Disclination Condensation: the red lines are disclination loops. When disclination proliferates and condenses, the bulk is saturated with close disclination loops. The end points of disclination line appears on the surface as disclination points, which is illustrated with purple dots.}
    \label{fig:loop}
\end{figure}

Now we restore the rotation symmetry by proliferating the $2\pi$ disclination as in Fig.\ref{fig:loop}. After proliferating and condensing the disclination loops, the rotation symmetry in the bulk is restored and the fermionic excitations is confined at low energy as fermion acquires $\pi$ Berry phase when going around the disclination loop. In sum, after the condensation of disclination loops in 8 copies of TSC, we obtained a short range entangled phase. We named such boson state as a $ topological$ $liquid$ $crystal$(TLC).

The disclination loop condensation procedure and the crystalline-liquid transition can be interpreted by the vortex-line condensation theory which is widely used in 3D superfluid-Mott transitions\cite{ye2015towards,ye2014vortex,vishwanath2013physics}. Start from the crystalline phase where the rotation symmetry is broken, the Goldstone mode of the nematic fluctuation is described by,
\begin{align} 
&\mathcal{L}=\frac{\kappa}{2}(\partial_{\mu} \theta^{[ab]})^2
\end{align}
$\theta^{[ab]}$ describes the Goldstone mode of the nematic order in the $a$-$b$ plane and index $\mu$ runs over spacetime. We made a Hubbard-Stratonovich transformation to the theory,
\begin{align} 
&\mathcal{L}=-\frac{1}{2\kappa} (J_{\mu}^{[ab]})^2+iJ_{\mu}^{[ab]}(\partial_{\mu} \theta^{[ab],s} +\partial_{\mu} \theta^{[ab],v})
\end{align}
$\theta^{[ab],s}$ and $\theta^{[ab],v}$ are the smooth and vortex part of the phase. Integrating out the smooth part gives the constrain $J_{\rho}^{[ab]}=\frac{1}{4\pi}\epsilon^{\rho \nu  \mu  \lambda} \partial_{\nu} B^{[ab]}_{\mu \lambda}$, $B_{\mu \lambda}$ is a two form antisymmetric gauge field. The effective action can therefore be written as,
\begin{align} 
&\mathcal{L}= \frac{i}{4\pi}\epsilon^{ \nu \rho\mu \lambda} \partial_{\nu} \omega^{[ab]}_{\rho} B^{[ab]}_{\mu \lambda} -\frac{h^{\mu\nu\lambda,[ab]} h_{\mu\nu\lambda,[ab]}}{48\pi^2 \kappa} \nonumber\\
&=\frac{1}{2} J^{dis, [ab]}_{\mu \lambda} B^{[ab]}_{\mu \lambda} -\frac{h^{\mu\nu\lambda,[ab]}h_{\mu\nu\lambda,[ab]}}{48\pi^2 \kappa} \nonumber\\
&J^{dis,[ab]}_{\mu \lambda} =\frac{i}{2\pi}\epsilon^{\mu \lambda \nu \rho} \partial_{\nu} \omega^{[ab]}_{\rho}=\frac{i}{2\pi}\epsilon^{\mu  \lambda \nu \rho} \partial_{\nu} \partial_{\rho}  \theta^{[ab]} \nonumber\\
&h^{\mu\nu\lambda,[ab]}=\partial_{\mu} B^{[ab]}_{\nu\lambda}+\partial_{\nu} B^{[ab]}_{\lambda\mu}+\partial_{\lambda} B^{[ab]}_{\mu\nu}
\end{align}
Here $\nu \rho\mu \lambda$ index runs over spacetime.
$J^{dis,[ab]}_{\mu \lambda} $ is the disclination current. This procedure inherits the boson-vortex duality method in 3+1d\cite{franz2007vortex}. As our disclination couples with the fermion in the TSC in terms of the spin-connection, integrating out the fermion gives rise to a geometry axion term,
\begin{align} 
&\epsilon^{\mu \nu \rho \lambda } \frac{1}{24\pi} \partial_{\mu}\omega^{[ab]}_{\nu} \partial_{\rho} \omega^{[ab]}_{ \lambda }
\end{align}
This axion term can be regarded as the disclination Witten effect in TSC where the monopole configuration of the disclination flux is bound to spin. To reflect this effect, we shall add a topological Berry phase term\cite{wang2015field,ye2015towards,ye2014vortex,vishwanath2013physics,kapustin2014coupling} for the two form gauge field $B$ which minimal couples with the disclination current. Consequently, the complete theory of the disclination loop condensation is,
\begin{align} 
&\mathcal{L}
=\frac{1}{2}J^{ab}_{\mu \lambda} B^{[ab]}_{\mu \lambda} -\frac{h^{\mu\nu\lambda,[ab]} h_{\mu\nu\lambda,[ab]}}{48\pi^2 \kappa}-\frac{3 \epsilon^{\mu \nu \rho \lambda}}{8\pi}B^{[ab]}_{\mu \nu}B^{[ab]}_{\rho \lambda}\nonumber\\
&\sim \frac{\kappa'}{2}(\partial_{[\mu} \Theta_{\lambda]}^{[ab]} +B^{[ab]}_{\mu \lambda})^2 -\frac{h^{\mu\nu\lambda,[ab]} h_{\mu\nu\lambda,[ab]}}{48\pi^2 \kappa}\nonumber\\
&-\frac{3\epsilon^{\mu \nu \rho \lambda}}{8\pi}B^{[ab]}_{\mu \nu}B^{[ab]}_{\rho \lambda}
\end{align}
The dynamics of  disclination loop is described by reparametrization invariant Nambu-Goto action. Once the disclination loop condensed, we arrive at the BF+BB type TQFT\cite{ye2015towards,ye2014vortex,vishwanath2013physics,wang2015field,gu2015multi},
\begin{align} 
\label{link}
&\mathcal{L}=- \epsilon^{\mu \nu \rho \lambda}\frac{3}{8\pi}B^{[ab]}_{\mu \nu }B^{[ab]}_{\rho \lambda}+\epsilon^{\mu \nu \rho \lambda}\frac{1}{4\pi} \partial_{\mu} \omega^{[ab]}_{\nu} B^{[ab]}_{\rho \lambda}
\end{align}
$B\wedge B$ is a topological term of the disclination loop configuration as long as U(1) and $\mathcal{T}$ is imposed. Integrating out $B$ field in Eq.\eqref{link}, one can obtain the wave function of the disclination condensed phase.
\begin{align} 
&|\Psi \rangle_{GS} \sim  \sum e^{i\epsilon^{ijk } \int dx^3 \frac{1}{24\pi} \omega^{[ab]}_{i} \partial_{j} \omega^{[ab]}_{k}}|\Psi[\omega] \rangle 
\end{align}
$\Psi[\omega]$ refers to all possible disclination loop configuration. The phase factor counts the number of links between two disclination loops and each extra linking contributes a phase factor  $\exp(i\frac{N_{link}2\pi}{6})$ to the wave function in the loop condensed phase. Resultantly, the wave function of the TLC after disclination condensation can be expressed as a superposition of all possible close disclination loop configurations. For each specific loop configuration, there is a coefficient $\exp(i\frac{N_{link}2\pi}{6})$ in the front which counts the number of linkings between every two loops\cite{xu2013wave,senthil2013integer}.

As a summary, we start from eight copies of $\mathcal{T}$ invariant TSC on a crystalline lattice which breaks rotation/translation symmetry. By proliferating the disclination loops, we restore the spatial rotation symmetry and meanwhile confine the fermions. The bulk theory we obtain is a topological liquid crystal(TLC) as a symmetry protected topological phase under $\mathcal{T}$ and spatial rotation symmetry. Such crystalline-liquid phase transition theory is characterized by the vortex loop condensation mechanism whose effective theory displays the BF+BB type TQFT.

\section{surface theory of TLC}
\label{surfaceTLC}
In this section, we devote to the surface state of 8 copies of TSC after disclination condensation. Even the 3D SPT phase themselves are gapped and short ranged entangled(SRE), the 2D surface of an SPT can be gapless, symmetry broken, or topological ordered. Different from the topological order in pure 2+1D, the surface topological order has obstruction where symmetry acts on the topological quasiparticle in an anomalous way\cite{chen2014anomalous}. In section \ref{tlc}, we show that loop condensation drives the bulk into a boson SRE state. However, on the surface, the end point of disclination loop(which is a disclination point) might have nontrivial braiding statistics which forbids itself from proliferation. 

In this part, we exam several possible surface states of the topological liquid crystal we obtained in section \ref{tlc}. We first look into the $\mathcal{T}$ broken surface state where the disclination has anyon statistics. Further, we look into the $\mathcal{T}$ symmetric gapped surface state. For even copies of TSC, one can always gap the surface state by pairing the surface Dirac cone and condense (multiple) superconducting vortices to restore the $\mathcal{T}$ symmetry\cite{metlitski2014interaction}.  Several pioneer work\cite{metlitski2014interaction,wang2014interacting,burnell2014exactly} had demonstrated that such vortex condensation can give rise to a rich class of surface topological order(STO) with obstruction. If we also restore the rotation and translation symmetry by disclination condensation in such $\mathcal{T}$ symmetric surface, the fermion sector would be confined but the other anyonic excitation remains unaffected. In addition, if we only restore the translation symmetry by condensation of disclination dipole(which is equivalent to an edge dislocation) on the surface, there exist exotic intertwined topological structure between the dislocation and quasiparticle.

\subsection{$\mathcal{T}$ broken surface}
\label{Tbroken}
The surface state of 8 copies of 3D topological superconductor contains 8 Majorana cones,
\begin{align} 
&H=\chi_i^T (i \partial_x \sigma_x+i \partial_y \sigma_z) \chi_i \nonumber\\
&\mathcal{T}: \chi^{\uparrow} \rightarrow ~ \chi^{\downarrow},~\chi^{\downarrow} \rightarrow  -\chi^{\uparrow}\nonumber\\
\end{align}
Assume we gap out the surface Majorana fermion by adding a mass term which breaks $\mathcal{T}$,
\begin{align} 
\label{mass}
&H=\Sigma_{i=1}^8 \chi_i^{T} (i \partial_x \sigma_x +i \partial_y \sigma_z+ m\sigma_y) \chi_i 
\end{align}
Couple the theory to the background vielbein field and integrating out the Majorana fermion, the effective theory of the disclination is,
\begin{align} 
&\mathcal{L}=\frac{1}{24\pi}\epsilon^{\mu\nu\rho} \omega^{[xy]}_{\mu} \partial_{\nu} \omega^{[xy]}_{\rho} 
\end{align}
The statistics of the disclination can be verified by the Hopf-term, 
\begin{align} 
&\mathcal{L}=\frac{1}{24\pi}\epsilon^{\mu\nu\rho} \omega^{[xy]}_{\mu} \partial_{\nu} \omega^{[xy]}_{\rho} \nonumber\\
& \rightarrow  \frac{\pi}{6} \epsilon^{\mu\nu\rho} J^{dis}_{\mu} \frac{\partial_{\nu}}{\partial^2}J^{dis}_{\rho}
\end{align}
$\frac{1}{2\pi}\epsilon^{\mu\nu\rho} \partial_{\nu} \omega^{[xy]}_{\mu}$ is the disclination current $J^{dis}_{\rho}$. The disclination has anyon statistics so a single $2\pi$ disclination cannot condensed at the surface.

\subsection{$\mathcal{T}$ invariant surface}
In this part, we investigate the $\mathcal{T}$ invariant surface theory of the topological liquid crystal protected by $\mathcal{T}$ and rotation symmetry. To obtain a $\mathcal{T}$ invariant gapped surface, we first follow the strategy raised by Metlitski $et~al.$\cite{metlitski2014interaction} by developing a superfluid order on the surface to gap the fermion (which meanwhile breaks $\mathcal{T}$) and finally restore the $\mathcal{T}$ by vortex condensation. To restore the spatial symmetry on the surface, we consider 2 possible situations. \textit{1}, the disclination on the 2D surface is proliferated so one can condense them to restore the rotation symmetry and meanwhile confine the fermion sector of the surface theory. \textit{2}, the disclination on the 2D surface is energetically confined, but the pair of disclination dipole(which is equivalent to a dislocation) can be proliferated to restore the translation symmetry.  At this stage, instead of developing a uniform superfluid state on the surface, we develop a Fulde-Ferrell(FF) state\cite{fulde1964superconductivity} where the superfluid phase factor is decorated with spatial modulation. The spatial dependent superfluid order is sensitive to dislocation defect and we observe interesting intertwined topological order between dislocation and superfluid vortex.

\subsubsection{restore the rotation symmetry}
\label{rotaionsymmetry}
Before we go into the detail of our result, we briefly review the theory developed by several pioneers\cite{metlitski2014interaction,wang2014interacting} where they obtained a surface topological order of 3D TSC without $\mathcal{T}$ breaking. As the topological liquid crystal we obtained in section \ref{tlc} is inherited from 8 copies of 3D TSC with strong interaction, we then have 8 copies of Majorana cone on the surface. One can always combine every two Majorana cone into a Dirac cone, then the surface theory is equivalent to 4 gapless Dirac theory,
\begin{align} 
&\Psi^{\uparrow}=\chi^{\uparrow}_1+i \chi^{\uparrow}_2,~\Psi^{\downarrow}=\chi^{\downarrow}_1+i \chi^{\downarrow}_2\nonumber\\
&H=\Sigma_{i=1}^4 \Psi^{\dagger}_i (p_x \sigma_x+p_y \sigma_z) \Psi_i   \nonumber\\
&\mathcal{T}: \Psi_{\uparrow} \rightarrow ~\Psi^{\dagger}_{\downarrow},~\Psi_{\downarrow} \rightarrow ~-\Psi^{\dagger}_{\uparrow}
\label{fermion}
\end{align}
Different from the Dirac theory in 2D, the fermion here is unusual under $\mathcal{T}$ transformation. The $\mathcal{T}$ operator acts on both the spin and the particle-hole channel. Now assume we gap the surface Dirac cones by turning on s-wave pairing of each cone. The condensation of cooper pair $O(\bm{r})_i=\epsilon_{\sigma \sigma'}  \Psi_{\sigma,i}(\bm{r})\Psi_{\sigma',i}(\bm{r})$ leads to a new superfluid term in the Hamiltonian,
\begin{align} 
&H=\Sigma_{i=1}^4 \Psi^{\dagger}_i (p_x \sigma_x+p_y \sigma_z) \Psi_i +\Delta^* O_i+\Delta O^{\dagger}_i  \nonumber\\
&\mathcal{T}: O(\bm{r})\rightarrow -O^{\dagger}(\bm{r})
\end{align}
The time reversal operation turns a superfluid creation operator into an annihilation operator. The superfluid order parameter breaks $\mathcal{T}$ but still invariant under the combination of time reversal and particle-hole $\mathcal{T}\times U_{\pi/2}$. Consequently, the vortex of the superfluid field is still a vortex under time reversal operation. Once we condense the vortex of the superfluid order, we would restore the $\mathcal{T}$ symmetry\cite{metlitski2014interaction}.
If we write the superfluid surface state back in the Majorana basis,
\begin{align} 
&H=\Sigma_{i=1}^4 \chi_i^{T} (p_x \sigma_x+p_y \sigma_z+\Delta_1 \sigma_y \tau_x+\Delta_2 \sigma_y \tau_z) \chi_i  
\end{align}
 As each vortex of the superfluid carries four Majorana zero modes, one has to condense the bound states of double vortices where the 8 Majorana can be gapped by F-K interaction without $\mathcal{T}$ broken. It was pointed out \cite{metlitski2014interaction,wang2014interacting} that condensation of double vortex give rise to $[e\mathcal{T}m\mathcal{T} ]\times [I,f]$ state, where $f$ is the original fermion which has trivial statistics with any other particle. After we proliferate the $2\pi$ disclination, the fermion $f$ (bogoliubov quasiparticle) is confined as it acquires a $\pi$ berry phase around the disclination and therefore the surface state becomes a pure $[e\mathcal{T}m\mathcal{T}]$ state.

To sum up, here we gap out the surface fermion and restore the $\mathcal{T}$ and rotation symmetry on the surface of the topological liquid crystal by disclination and vortex condensation. The fermion on the surface is confined by the disclination and the surface topological order is the $[e\mathcal{T}m\mathcal{T}]$ state.

\subsubsection{surface superfluid with Fulde-Ferrell(FF) order}
In this section, we move to another condition where the surface disclination is energetically unfavored and therefore confined but the dislocation is deconfined. In crystalline phases, a dislocation can be fused by a pair of disclination dipole with opposite charge. The two disclinations of opposite charge have attractive interaction at 2D and thus being confined as a bound state of disclination dipole at low temperature. When the temperature goes up, the disclination is deconfined and proliferated which give rise to an isotropic liquid by lattice melting(which is the situation we studied in section ~\ref{rotaionsymmetry}). A pair of disclination dipole forms a dislocation in the crystal phase. By proliferating such dislocations, the translation symmetry is restored and the system goes into a nematic surface state. This is the case we would focus on in the following paragraph. 

To acquire an intertwined topological order between dislocations and surface topological quasiparticles, we develop a Fulde-Ferrell(FF) superfluid order $\Delta= \Delta_{\bm{Q}} e^{i\bm{Qr}}$ instead of the uniform superfluid order. The phase factor is dressed with an amplitude oscillation in space along the FF wave vector $\bm{Q}$. After dressing the superfluid order with $e^{i\bm{Qr}}$, the pairing term becomes sensitive to lattice defect. $e.g$, If we have a dislocation in the FF order as $\epsilon^{ij} \partial_i \partial_j \bm{Qr}= 2\pi$, the superfluid boson $\Delta$ has a phase winding $2\pi$ around the dislocation. The FF superfluid order consequently intertwined the dislocation(defect of translation) with the vortex (defect of the superfluid order).
\begin{figure}[h]
    \centering
    \includegraphics[width=0.4\textwidth]{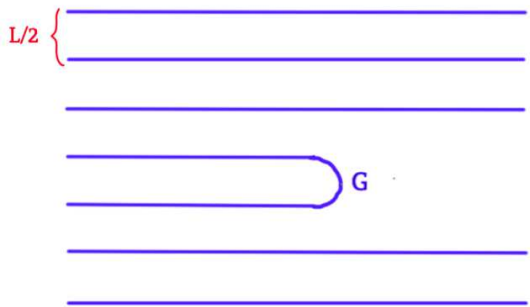}
    \caption{FF order of superfluid: The blue lines displays the area where $Qr=n \pi$. The spacing between two lines is half of the periodicity $L$. At point G, there is a $2\pi$ dislocation of the FF order and the pairing field has a $2\pi$ winding around the dislocation.}
\end{figure}

The FF superfluid surface state breaks both $\mathcal{T}$ and translation symmetry. Our goal is to study the $\mathcal{T}$ and translation invariant surface state of the topological liquid crystal. In order to restore the symmetry, we need to condense the superfluid vortex and melt the soft lattice via dislocation condensation. The dislocation condensation, driven by quantum melting of the soft lattice, is distinguished from the conventional classical melting transition induced by thermal fluctuation. The term `soft lattice' refers to a spatial dependent order parameter formed by electrons or quasiparticles($e.g.$ the FF superfluid order) rather than the classical ionic lattices. Ergo, the dislocation of the soft lattice is the defect of the spatial dependent order parameter and the condensation of the dislocation can be approached in terms of Ginzburg-Landau type theory akin to the vortex condensate in Superfluid-Mott transitions.

However, due to the topological nature of the surface state, the vortex and dislocation of the FF order parameter might carry Majorana zero mode and convey fractional braiding statistics. We have to investigate the character of the symmetry defect carefully before we condensed them. In the next paragraphs, we would reveal that either a single $2\pi$ dislocation or a superfluid vortex carries Kramers doublet. In order to obtain a $\mathcal{T}$ invariant gapped surface state, our strategy is to condense the bound state of a dislocation and a superfluid vortex. Such bound state condensation destroys the surface superfluid FF order and meanwhile restores the translation and $\mathcal{T}$ symmetry.

\subsubsection{dislocation theory}
\label{dislocationtheory}
The superfluid order parameter is written as $\Delta= |\Delta_{\bm{Q}}| e^{i\bm{Qr}+\phi} e^{i\Theta}$, $\Theta$ is the vortex of the superfluid and $\phi$ is the phase fluctuating of the FF order. Both a vortex in $\Theta$ and a $2\pi$ dislocation would give rise to a $2\pi$ winding number of the superfluid boson. The $2\pi$ winding of the $\Delta$ field traps Majorana zero mode for every 2 copies of the TSC surface state. Since we have 8 copy of the TSC surface state, we have in total 4 Majorana zeros modes trapped inside the $2\pi$ dislocation. These 4 Majorana zeros modes cannot be gapped by any $\mathcal{T}$ invariant  interaction. Resultantly, one cannot obtain a $\mathcal{T}$ invariant  gapped state by single vortex condensation.

To investigate the statistics of the dislocation we follow the argument made by Metlitski $et~ al.$\cite{metlitski2014interaction}. The complex fermion we defined in Eq. \eqref{fermion} has U(1) gauge symmetry and the complex fermion current therefore couples with the U(1) gauge field $a$. Accordingly, the superfluid boson composed of fermion pairs minimal couples with the U(1) gauge field with charge 2.  A $2\pi$ dislocation in the FF wave vector give rise to a $2\pi$ winding number of the superfluid boson, the dislocation would trap a $\pi$ gauge flux(of $a$) in order to get rid of the logarithmic divergence from the kinetic energy of dislocation $(\partial_{\mu} \phi -2a)^2|\Delta|^2 $. The $\pi$ gauge flux is confined inside the dislocation on the top surface(superfluid state) and leaks down to the bulk and the bottom surface.

Now imagine the top surface is in the FF superfluid state while the bottom surface breaks time reversal as in section \ref{Tbroken}. Since the domain wall between the superfluid state and $\mathcal{T}$ broken state has central charge $c=2$, the whole slab can be regarded as 4 copies of 2D $p+ip$ superconductor. The dislocation only lives on the top surface, but the $\pi$ gauge flux it traps penetrates the whole slab. The anyon theory of the bottom surface and the whole slab is affected by the flux emanating from the dislocation on the top.

The anyon theory of the flux excitation in 2D $p+ip$ superconductor is well studied\cite{kitaev2006anyons}. For 4 copies of $p+ip$, the $\pi$ flux trapped inside the vortex excitation is a semion(we label it as the $e$ particle). If we fuse this semion with a bogoliubov quasiparticle, we obtain the $m$ particle which is also a semion and meanwhile $e$ and $m$ are mutually bosons. 

The flux trapped by the dislocation is a two semion theory $U(1)_2\times U(1)_2$. As the bulk of the slab itself does not have any topological order, we only need to consider the anyon theory in the top and bottom surface and the sum of them should be equivalent to the $U(1)_2\times  U(1)_2$ theory\cite{metlitski2014interaction}. The bottom surface has four copies of massive Dirac cone which breaks $\mathcal{T}$. The flux on the bottom surface has a $U(1)_{-2}$ theory. Accordingly, the top surface is a $U(1)_2\times U(1)_2 \times U(1)_{-2}$ theory where the $e$ , $m$ particle are mutually semions and self-bosons. The four Majorana zero modes inside the $e$,$m$ particle can be split into two degenerate doublets with opposite fermion number parity. Resultantly, both $e$ and $m$ carries a Kramers doublet.  To avoid ambiguity in our later discussion, we label them as $e^{dis}$,$m^{dis}$.  As the single dislocations($e^{dis}$,$m^{dis}$) carries Kramers doublet, $\mathcal{T}$ symmetry prohibits them from condensation.

\subsubsection{vortex theory}
\label{doubleetmt}
For the FF state superfluid order, a $2\pi$ vortex of the superfluid boson also traps $\pi$ gauge flux who confines inside the dislocation on the top surface(superfluid state) and leaks down to the bulk and the bottom surface. To investigate the statistics of the surface vortex we follow the argument made by Metlitski $et~ al.$\cite{metlitski2014interaction}. The vortex of the superfluid field, label as $e^{v}$, has mutual semion statistics with $m^{v}$ which is the bound state of a vortex and bogoliubov quasiparticle. Meanwhile, both $e^{v},m^{v}$ are self bosons.

Now we have in total 4 types of topological quasiparticle $e^{dis}$,$m^{dis}$ and $e^{v}$,$m^{v}$ and it is essential to figure out their mutual statistics. Following the same slab argument we did in section \ref{rotaionsymmetry}, both a vortex of the superfluid and a $2\pi$ dislocation in the FF wave vector traps $\pi$ flux penetrating along the slab. The slab is equivalent to 4 copies of $p+ip$ superconductor so the flux acquires a phase of $\pi$ when winding around the other. The dislocation and the vortex in the whole slab are mutually semions. Take away the mutual statistical phase contribution from the bottom surface where $\pi$ flux has a $U(1)_{-2}$ theory, the $e^{dis}$ and $e^{v}$ on the top surface are mutual bosons. The same argument applies for $m^{dis}$ and $m^{v}$ which are also mutually bosons ,while $e^{dis}$ and $m^{v}$ (also $m^{dis}$ and $e^{v}$) are mutual semions. All of these topological quasiparticles carries Kramers doublet whose condensation would still break $\mathcal{T}$ symmetry. However, the bound state between a $2\pi$ vortex($e^{v}$) and a $2\pi$ dislocation($e^{dis}$) carries in total eight Majorana zero modes which could be gapped by F-K type interaction. The vortex-dislocation bound state is therefore a Kramers singlet whose condensation restores both $\mathcal{T}$ and translation symmetry.
After vortex-dislocation bound state condensation, the surface theory is a $\mathcal{T}$ invariant topological nematic state. The remaining $e^{dis}$,$m^{dis}$ and $e^{v}$,$m^{v}$ are deconfined quasiparticles with nontrivial statistics.The surface topological order is a double $[e\mathcal{T}m\mathcal{T}]$ state which cannot be realized in a pure 2D system.
The K matrix formalism of the anyonic theory on the surface is,
\begin{align}
&\mathcal{L}= K_{IJ} \epsilon^{\mu \nu \rho} a_I^{\mu}  \partial_{\nu} a^{\rho}_J+a^{\mu}_1(J^{\mu}_{e^{v}}+J^{\mu}_{e^{dis}})+a^{\mu}_2(J^{\mu}_{m^{v}}+J^{\mu}_{m^{dis}})
\nonumber\\
&K=\frac{1}{4\pi}
\begin{pmatrix} 
0 & 2 \\
2 & 0\\
\end{pmatrix}
\end{align}
$J^{\mu}_{a}$ refers to different topological quasiparticle current. Integrating out the auxiliary gauge field $a$ reveals the self and mutual braiding statistics we obtained in the previous paragraph. 

\subsubsection{Geometry aspects on SPT classification of interacting 3D $\mathcal{T}$ TSC}
Earlier work had pointed out \cite{wang2011topological,ryu2012electromagnetic} that the geometry response in 3D $\mathcal{T}$ TSC only shows a $Z_2$ class, which is weaker than the integer classification of the non-interacting system or the $Z_{16}$ class at the interacting level.  It was suspected that the geometry response cannot fully capture or identify different class of phase in 3D TSC. However, based on our conclusion in section \ref{dislocationtheory}, the FF superfluid order on the surface intertwines the dislocation with the superfluid vortex and condensation of these two types of defect give rise to new topological order. Depending on the number of copies we have for TSC, the dislocations exhibit different braiding statistics at the surface and this provides us a new way to identified different SPT phases from geometry degree of freedom.

\subsection{NL$\sigma$M  on the surface of TSC} \label{sssec:Nlsm}
The surface topological order in the $Z_{16}$ class of 3D TSC can be reached in a variety of ways. Beside the vortex condensation argument, the surface topological order for eight copies of 3D TSC can also be obtained via NL$\sigma$M theory with a topological theta term\cite{bi2014bridging,you2014symmetry}. Assume we couple 8 copies of 3D TSC with an O(5) vector $\bm{n}$, write the theory in the Majorana basis,
\begin{widetext} 
\begin{align} 
&\mathcal{H}=\sum_{\bm{k}} ~\chi_{\bm{k}}^T (-k_x \sigma^{33000}-k_y\sigma^{10000}-k_z \sigma^{31000}+m \sigma^{20000}+n_1 \sigma^{32212}+n_2 \sigma^{32232}+n_3 \sigma^{32220}+n_4 \sigma^{32300}+n_5 \sigma^{32100}) \chi_{\bm{-k}} \nonumber\\
&\mathcal{T}: \bm{n} \rightarrow ~- \bm{n}, \chi  \rightarrow \mathcal{K}i\sigma^{32000} \chi
\end{align}
\end{widetext} 
Each component of $\bm{n}$ couples with a fermion bilinear. The fermion degree of freedom could be confined by the $Z_2$ flux and the remaining theory for the bosonic degree of freedom($\bm{n}$ vector field) is obtained by integrating out the fermion. The effective bulk theory of the O(5) boson vector is described by O(5) NL$\sigma$M with  topological theta term for $\Theta=2\pi$\cite{bi2014bridging,you2014symmetry}.
The surface corresponding Hamiltonian is,
\begin{widetext} 
\begin{align} 
\label{surface}
&\mathcal{H}=\sum_{\bm{k}} ~\chi_{\bm{k}}^T (-k_x \sigma^{3000}-k_y\sigma^{1300}+n_1 \sigma^{1112}+n_2 \sigma^{1132}+n_3 \sigma^{1120}+n_4 \sigma^{1200}+n_5 \sigma^{2000}) \chi_{\bm{k}} \nonumber\\
&\mathcal{T}:  \chi  \rightarrow \mathcal{K}i\sigma^{2300} \chi
\end{align}
\end{widetext} 
This surface state as an interface which separates 3D NL$\sigma$M from $\Theta=2\pi$ to $\Theta=0$ can be described 
by the O(5) WZW theory with $k=1$\cite{you2014symmetry,xu2013nonperturbative}. If we break O(5) to O(4) by taking the $n_5=0$ limit, the surface theory is therefore described by O(4) NL$\sigma$M with $\Theta=\pi$.
\begin{align} 
&\mathcal{L}=(\partial_{\mu} \bm{n})^2+\frac{i\pi}{\Omega_3} \epsilon^{ijkl} n^i \partial_x n^j \partial_y n^k \partial_0 n^l
\end{align}
$\Omega_3$ is the is the volume of $S^3$. The O(4) NL$\sigma$M with $\Theta=\pi$ is critical when we have the exact O(4) symmetry. But once we break the symmetry to $U(1)\times U(1)$(with respect to the O(2) rotor for $n_1,n_2$ and $n_3,n_4$), the theory is gapped and topological ordered.
Imagine we have a vortex $\Theta^{12}$ between $n_1,n_2$ and vortex $\Theta^{34}$ between $n_3,n_4$, the $\Theta$ term shows that these two vortices have mutual semion statistics, which agrees with the vortex condensation argument.

Now we dressed the O(4) vector with some spatial modulation vector. Define
 \begin{align} 
&N_1=n_1+in_2=|N_1| e^{i\Theta^{12}} e^{i\bm{Qr}},\nonumber\\
&N_2=n_3+in_4=|N_2| e^{i\Theta^{34}} e^{i\bm{\bar{Q}\bar{r}}},\nonumber\\
& \bar{Q} \bot Q 
\end{align}
If we regard $N_1,N_2$ as a classical O(2) rotor, the phase modulation $e^{iQr}$ therefore generates a spiral order parameter which transmutes the polar angle of the rotor with a uniform phase shift along the $\bm{r}$ direction.

If there is a $2\pi$ dislocation along the $r$(or $\bar{r}$) direction, the vector $N_i$ experience phase winding of $2\pi$ when going around the dislocation point. Now we define $Qr=\alpha^{dis},\bar{Q}\bar{r}=\beta^{dis}$. Integrating out the fermion leads to,
\begin{align} 
&\mathcal{L}=\frac{i\pi}{\Omega_3} \epsilon^{ijkl}  (\partial_x (\Theta^{12}+\alpha^{dis})) \partial_y  (\partial_0 (\Theta^{34}+\beta^{dis}))
\end{align}
The vortex $\Theta^{12}$ and dislocation $\alpha^{dis}$ has mutual semion statistics with $\Theta^{34}$ and $\beta^{dis}$. This agrees with our former result on vortex condensation where $e^{dis}, e^{v}$ has mutual semion statistics with $m^{dis}, m^{v}$. 
From Eq. \eqref{surface}, either a vortex or a $2\pi$ dislocation of $N_i$ would trap 4 Majorana zero mode on the surface. Following the similar argument as You $et~al.$\cite{you2014symmetry}, the dislocation carries a Kramers doublet. To restore the translation symmetry of the surface without gap closing, one needs to condense at least double dislocation to get rid of the redundant zero modes. After double dislocation condensation, the surface state is in a nematic phase with $Z_2$ topological order where two dislocation with Burgers vector in perpendicular direction has mutual semion statistics. 

Here, it is essential to point out that the fermion bilinears coupled with $n_1,n_2$ in Eq. \eqref{surface} is precisely the surface superfluid order we discussed in Section \ref{doubleetmt}. When the O(5) rotor is disordered, the surface superfluid order is destroyed and the vortex between $n_1,n_2$, as a deconfined excitation, is exactly the $2\pi$ vortex of the superfluid($e^v$) we discussed in Section \ref{dislocationtheory} who carries a Kramers doublet.  Meanwhile, the vortex of $\Theta^{34}$ who has mutual semion statistics is actually the $m^v$  in Section \ref{doubleetmt}.

To conclude, in this section, we retest the surface topological order of the topological liquid crystal inherited from 8 copies of 3D TSC via the NL$\sigma$M argument. By dressing the rotor vector with a spiral order, we obtained a consistent topological nematic surface as we did in section \ref{dislocationtheory}.

\section{topological quantity of dislocation in NL$\sigma$M and Levin-Gu model in 2d} \label{gu}

\subsection{dislocations in 2D BSPT described by NL$\sigma$M models}
\label{bspt}

Motivated by the result in Section \ref{sssec:Nlsm}, we first look into the interplay between crystalline defect(dislocation) and symmetry flux in a 2D SPT whose low energy theory is inherited from NL$\sigma$M with topological $\Theta$ term. The topological $\Theta$ term in NL$\sigma$M provides a straightforward way to classify and characterize boson SPT in 2D\cite{xu2013wave,levin2012braiding,wang2015field}. The gapless(or symmetry breaking) edge state of the SPT is encoded in the $\Theta$ term who reduced to a Wess-Zumino-Witten term on the edge. In addition, after symmetry gauging, the nontrivial braiding statistics of the symmetry flux can also be verified via the $\Theta$ term.

To obtain a topological NL$\sigma$M with $\Theta$ term, we start with 4 copies of $p \pm ip$ superconductor and couple the fermions with an O(4) vector, whose effective coupling can be simplified as\cite{you2014symmetry},
\begin{widetext} 
\begin{align} 
&\mathcal{H}=\sum_{\bm{k}} ~\chi_{\bm{k}}^T (-k_x \sigma^{3000}-k_y\sigma^{1000}+m\sigma^{2300}+n_1 \sigma^{2100}+n_2 \sigma^{2212}+n_3 \sigma^{2220}+n_4 \sigma^{2232}) \chi_{\bm{-k}} 
\end{align}
\label{2dtheta}
\end{widetext} 
This theory can be viewed as a descendant of Eq. \eqref{surface} which describes the surface theory of 3D TSC. If we reduce the O(5) vector to O(4) in Eq. \eqref{surface} by assigning a nonzero expectation value for $n_5=m$, then Eq. \eqref{surface} becomes equivalent to the theory we are investigating here.
In the O(4) disordered phase, the effective theory is a boson NL$\sigma$M at $\Theta=2\pi$\cite{you2014symmetry}. 

\begin{align} 
&\mathcal{L}=(\partial_{\mu} \bm{n})^2+\frac{i2\pi}{\Omega_3} \epsilon^{ijkl} n^i \partial_x n^j \partial_y n^k \partial_0 n^l
\end{align}

If the fermion gap is large, the fermion degree of freedom is absent at low energy spectrum so the system can be regarded as a pure boson SPT\cite{you2014symmetry}.
Dress each O(2) rotor with spiral order as we did in section \ref{sssec:Nlsm},
\begin{align} 
&N_1=n_1+in_2=|N_1| e^{i\Theta^{12}} e^{i\bm{Qr}}, \nonumber\\
&N_2=n_3+in_4=|N_2| e^{i\Theta^{34}} e^{i\bm{\bar{Q}\bar{r}}}, (\bar{Q} \bot Q) \nonumber\\
&Z_2^A: n_{1},n_2\rightarrow -n_1,-n_2; Z_2^B: n_{3},n_4\rightarrow -n_3,-n_4
\end{align}

The $N_i$ rotors are dressed with a spiral order which enable the rotor to oscillate along $\bm{r}$ direction. After integrating out the fermion, one obtains a 2D boson SPT state described by O(4) NL$\sigma$M with topological $\Theta$ term protected by $Z^A_2 \times Z_2^B$ symmetry. The theta term of the O(4) NL$\sigma$M with spiral order is,
\begin{align} 
\label{gauge}
&\mathcal{L}=\frac{i2\pi}{\Omega_3} \epsilon^{ijkl}  (\partial_x (\Theta^{12}+\alpha^{dis})) \partial_y  (\partial_0 (\Theta^{34}+\beta^{dis}))
\end{align}
The theory itself is an SPT state so there is no anyonic statistics. However, it was pointed out \cite{xu2013wave,levin2012braiding,wang2015field} that if we gauge the symmetry of SPT phase, the gauge flux can exhibit topological order. If we gauge the $Z^A_2 \times Z_2^B$ symmetry and couple the theory with the vision excitation carrying a $\pi$ flux, the vison line of $Z^A_2$($Z_2^B$) is bound with a $\pi$ dislocation for $\alpha^{dis}$ ($\beta^{dis}$) or half vortex of $\Theta^{12}$($\Theta^{34}$). According to the topological theta term in Eq. \eqref{gauge}, the vison braiding procedure is equivalent to braiding a half-dislocation of $\alpha^{dis}$(or half vortex of $\Theta^{12}$) with a half-dislocation of $\beta^{dis}$(or half vortex of $\Theta^{34}$). The braiding process finally gives $\pi/2$ Berry phase, indicates the nontrivial braiding statistics between the half dislocation/vortex.

To summarize, if we decorated the rotors in the 2D boson SPT with spiral order and gauge the $Z^A_2 \times Z_2^B$ symmetry, the dislocation and the gauge flux show intertwined topological ordered. If we further break the symmetry from $Z^A_2 \times Z_2^B$ to a global $Z_2$ where $\vec{n}\rightarrow -\vec{n}$ under the $Z_2$ symmetry transformation,  the vison of the $Z_2$ gauge field is a bound state between the two vision for $Z^A_2, Z_2^B$. The braiding between two vision becomes the self-rotation of the bound state. Since the Berry phase accumulated by the self-rotation is $\pi/2$, the two half-dislocations bounded with the vison is a self-semion.

\subsection{Stripy Levin-Gu model}
Based on the study in section~\ref{bspt}, the 2D SPT described by an O(4) NL$\sigma$M with spiral order can exhibit exotic braiding statistics between the dislocation and gauge flux. It is well known that an O(4) NL$\sigma$M with topological theta term for $\Theta=2\pi$ is equivalent to the Levin-Gu model if we take the Ising limit of O(4) vector\cite{xu2013wave,levin2012braiding}. The original O(4) vector can be reduced to $Z_2 \times O(3)$ as $(\phi,\bm{\vec{n}})$ and assume system energetically favors vector $\bm{\vec{n}}$ over $\phi$, the wave function can be therefore written as,
\begin{align} 
& |\Psi \rangle=\int D\vec{n} ~\exp(\frac{i\pi}{8\pi} \int dx^2 \epsilon^{ij} \epsilon^{abc} n_a \partial_i n_b \partial_i n_c) |\vec{n}(\vec{x}) \rangle \nonumber\\
&=\sum_{\vec{n}} (-1)^{N_s} |\vec{n}(\vec{x})\rangle
\end{align}
The wave function runs over a coherent state of all superposition of $\vec{n}(\vec{x})$ configuration and each configuration is dressed with a sign structure depending on the skyrmion number($N_s$) in the configuration.

Now assume we further break O(3) down to $Z_2$ by  energetically favor $n_1$ over other component. Then the skyrmion number reduced to the domain wall number($N_d$) of $n_1$ and the wave function reduced to,
\begin{align} 
& |\Psi \rangle=\sum_{n_z} (-1)^{N_d} n_1(\vec{x})
\end{align}
Which is exactly the Levin-Gu wave function\cite{xu2013wave,levin2012braiding}.

In our previous discussion, it turns out that the O(4) NL$\sigma$M with spiral order decoration exhibit nontrivial braiding statistics between flux and dislocation after we gauge the symmetry.  When we break the O(4) symmetry of this model to $Z_2$ limit and gauge the $Z_2$ symmetry, the flux exhibit semion/antisemion statistics which is studied by several early pioneers\cite{xu2013wave,levin2012braiding}. In this section, we decorate the Levin-Gu model with a stripe modulation. Such decoration, similar to what we did in section \ref{bspt}, intertwines the spatial translation defect(dislocation) with the $Z_2$ flux. It turns out that after we gauge the symmetry, the dislocation and the $Z_2$ flux have semion/antisemion statistics.
\begin{figure}[h]
    \centering
    \includegraphics[width=0.35\textwidth]{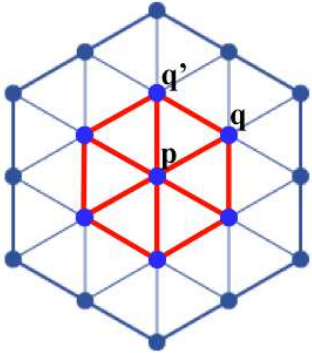}
    \caption{Levin-Gu model on the triangle lattice}
\end{figure}
\begin{figure}[h]
    \centering
    \includegraphics[width=0.35\textwidth]{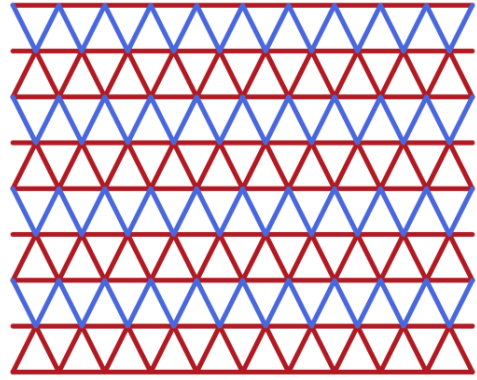}
    \caption{Levin-Gu model with some spatial modulation.  The sign before $\sigma^z_q \mu^z_{qq'}  \sigma^z_{q'}$ varies in space, depending on whether the bond lives on the even/odd stripe on the lattice. All the blue bonds have plus sign while the red bonds have minus sign for $\sigma^z_q \mu^z_{qq'}  \sigma^z_{q'}$. 
}
    \label{LG}
\end{figure}

To start with, we would decorate the Levin-Gu model with a stripy modulation, which enables the reciprocation between dislocation and symmetry flux.
The stripy Levin-Gu model coupled with $Z_2$ gauge flux is,
\begin{align} 
&\mathcal{H}=-\Sigma_p B_p O_p-\Sigma_{pqr} \mu^z_{pq} \mu^z_{pr} \mu^z_{qr}\nonumber\\
&B_p=-\sigma^x_p \prod_{pqq'} i^{\frac{1\pm \sigma^z_q \mu^z_{qq'}  \sigma^z_{q'}}{2}} ,O_p=\prod_{pqr} (1+\mu^z_{pq} \mu^z_{pr} \mu^z_{qr})/2
\end{align}

The $\sigma$ operator acting on the Ising spin variable lying on the site of the triangle lattice. The $\mu$ operator is the $Z_2$ gauge flux lying on the bond connecting nearest sites. $\Sigma_{pqr} \mu^z_{pq} \mu^z_{pr} \mu^z_{qr}$ term ensures the ground state(GS) is flux free and $i^{\frac{1\pm \sigma^z_q \mu^z_{qq'}  \sigma^z_{q'}}{2}} $ is the minimal coupling between Ising variable and $Z_2$ gauge flux. Different from the original Levin-Gu model, the $Z_2$ charge carried by the Ising variable is spatial dependent due to the stripe modulation of our model. As a result, the sign before $\sigma^z_q \mu^z_{qq'}  \sigma^z_{q'}$ varies in space, depending on whether the bond lives on the even/odd stripe on the lattice. The horizontal bond is always negative. The sign structure is displayed in Fig.\ref{LG}, all the blue bonds have plus sign while the red bonds have minus sign for $\sigma^z_q \mu^z_{qq'}  \sigma^z_{q'}$. 

\begin{figure}[h]
    \centering
    \includegraphics[width=0.3\textwidth]{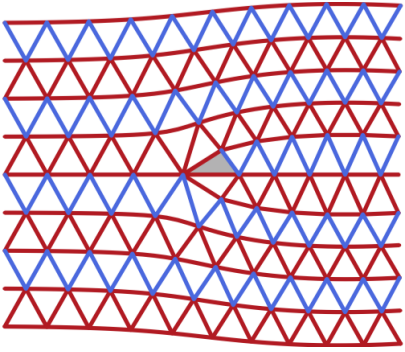}
    \caption{Levin-Gu model with dislocation. The gray triangle near the dislocation carries a flux which cannot be gauge away locally.}
        \label{LGD}
\end{figure}

The GS wave function of this stripy modulation Levin-Gu model is still a superposition of the all domain wall configurations (for the Ising sector)where each domain wall configuration has a sign in the front depending on the number of domain walls. For the $Z_2$ gauge sector, all the flux in the triangle should be flux free so the gauge choice can be $\mu^z_{aa'}=1$ for gauge field on the red bond while $\mu^z_{aa'}=-1$ for gauge field on the blue bond.
Thus, we write the wave function as,

\begin{align} 
\label{bond}
|\Psi>=(\sum_{m} (-1)^{N_{d}} |{m}\rangle )\prod_{red} | (\mu_z=1)\rangle  \prod_{blue} |(\mu_z=-1) \rangle 
\end{align}
${m}$ refers to different Ising spin configuration.
Now, we consider the spatial modulation Levin-Gu model on the triangle lattice in the presence of dislocation. The Hamiltonian remains the same and the sign in front of $\sigma^z_q \mu^z_{qq'}  \sigma^z_{q'}$ depends on the location of the bond, which is illustrated by color in Fig.\ref{LGD} as a dislocation lattice.

All the terms in the Hamiltonian still commute each other. However, if we still choose the gauge sector according to the bond color in Eq.\eqref{bond}, the triangle with blue bond near the dislocation point carries a flux which cannot be gauge away locally. To get the GS wave function, we need to eliminate the flux. The GS wave function of this stripy Levin-Gu model with dislocation is,
\begin{widetext}
\begin{align} 
\label{GS}
&|\Psi>=V_{a} (\sum_{m} (-1)^{N_{dm}} |{m}\rangle)  \prod_{red} |(\mu_z=1) \rangle \prod_{blue} |(\mu_z=-1)\rangle  \nonumber\\
&V_{a}=\prod_{<pq>\bot a} ~\mu^x_{pq} ~\prod_{<pqq'>,r} ~ i^{\frac{1\pm \sigma^z_q \mu^z_{qq'}  \sigma^z_{q'}}{2}}  ~\prod_{<pqq'>,l} ~(-1)^{s_{pqq'}} \prod_{<pqq'> \in a} (1+\mu^z_{pq} \mu^z_{pr} \mu^z_{qr})/2\nonumber\\
& s_{pqq'}=\frac{(1\pm \sigma^z_p \mu^z_{pq}  \sigma^z_{q})(1\pm (-1)\sigma^z_p \mu^z_{pq'}  \sigma^z_{q'} )}{4}
\end{align}
\end{widetext}
Here $V_{a}$ is a flux generating operator starting from the triangle near the dislocation as is illustrated in Fig.\ref{LGD}. The $V_{a}$ operator creates/annihilates a $Z_2$ flux at the end of the string. In this way, the flux carried by the dislocation is therefore annihilated by the flux string. 

It was pointed out\cite{levin2012braiding} that due to the commutation relation between string operator $V_{a}V_{b}=-V_{b}V_{a}$, braiding procedure between two gauge flux gives a minus sign. As is explicit from the ground state wave function, in the presence of dislocation in Eq.\eqref{GS}, a dislocation is accompanied with a string operator creating $Z_2$  flux in the dislocation point. Winding a flux line along the dislocation would attribute a $\pi$ Berry phase, the same berry phase also appears when we braiding two dislocations. Therefore, the dislocation is a self-semion and it has semion statistics with the gauge flux. This agrees with our field theory result in Eq.\eqref{gauge}.

\section{conclusion and outlook}
In this paper we studied the novel properties of geometry defects in SPT phases. We argue that the condensation of disclination in 8 copies of 3D $\mathcal{T}$ TSC would confine the fermion and afterward turn the system into a 3D boson topological liquid crystal(TCL), whose bulk is SRE but the surface displays topological order. In addition, we investigate the interplay between dislocation and superfluid vortex on the surface of TCL via FF surface superfluid order. It turns out that after condensation of double superfluid vortex and double dislocation, we obtain a $\mathcal{T}$  and translation invariant surface topological order, a double $[e\mathcal{T}m\mathcal{T}]$ state, which cannot be realized in pure 2D system.
We also look into the dislocation theory in 2D boson SPT described by the $O(4)$ NL$\sigma M$ with a $\Theta$ term(when $\Theta=2\pi$). After we dress the $O(4)$ vector with some spiral order and gauge the symmetry, the dislocation has nontrivial braiding statistics with the gauge flux. Further reduced the $O(4)$ vector to Ising limit, we arrive at the Levin-Gu model with stripy modulation. If such stripy Levin-Gu model is embedded in a lattice with dislocation, the dislocation has semion statistics with the $Z_2$ gauge flux.

\begin{acknowledgements}
We are grateful to Cenke Xu, Chetan Nayak, Peng Ye, Andreas Ludwig,Thomas Scaffidi, Tarun Grover and Zhen Bi for insightful comments and discussions. This work was supported in part by the National Science Foundation through grants DMR-1408713 (YY) at the University of Illinois, grants PHY11-25915(YY,YZY), DMR-1151208 (YZY) at The Kavli Institute for Theoretical Physics and David and Lucile Packard Foundation(YZY). YY acknowledges the KITP Graduate fellowship program, where this work was initiated.
\end{acknowledgements}

\providecommand{\noopsort}[1]{}\providecommand{\singleletter}[1]{#1}%

\end{document}